\documentclass[aps,pra,reprint]{revtex4-2}
\usepackage{amsmath,amssymb,bm}
\usepackage[colorlinks=true]{hyperref} 
\usepackage{comment}

\begin{document}

\title{Comment on ``Angular momentum dynamics of vortex particles in accelerators''}

\author{S.S.~Baturin}
\email{s.s.baturin@gmail.com}
\affiliation{School of Physics and Engineering, ITMO University, St.\ Petersburg 197101, Russia}

\begin{abstract}
We comment on Ref.~\cite{Karlovets}, which proposes a BMT-like equation for the mean kinetic orbital angular momentum (OAM) of vortex particles in accelerator fields and draws spin-like conclusions about depolarization, resonances, and control. We show that the proposed closure is not generally valid even at the mean-value level. In the authors' own homogeneous-field model, Eq.~(8) already makes $\langle L_z\rangle$ depend on the packet second moment $\langle \rho^2\rangle(\tau)$; for an exact family of breathing Landau/LG packets this yields an explicit oscillation incompatible with Eq.~(9) except in the nongeneric matched case. Moreover, the Appendix A assumption that mixed correlators are negligible suppresses the transverse kinetic-OAM components themselves, since those correlators are precisely the building blocks of $L_x$ and $L_y$. We also stress that, even if a closed equation for $\langle \hat{\mathbf L}\rangle$ were available, it would still not constitute a transport equation for a vortex quantum state. Mean-OAM transport does not determine OAM spectra, inter-mode coherences, or fidelity. State-level claims therefore require a mode-resolved density-matrix treatment rather than an Ehrenfest equation for a low-order moment.
\end{abstract}

\maketitle

\section{Introduction}

Reference~\cite{Karlovets} studies the dynamics of vortex particles in accelerator fields and advances two logically distinct claims: first, that the mean kinetic OAM obeys a closed BMT-like precession law, Eq.~(9); second, that this supports state-level conclusions phrased in the language of polarization, depolarization, resonances, and spin-like control. In this note we argue that both steps are problematic. At the mean-value level, the proposed closure is not generally valid. The authors' own Eq.~(8) shows that $\langle L_z\rangle$ depends on second moments of the packet, and the Appendix A truncation used to remove these terms suppresses the transverse kinetic-OAM itself. At the state level, even a correct closed equation for $\langle \hat{\mathbf L}\rangle$ would still not amount to a transport theory for a twisted quantum state, because a few low-order moments do not determine OAM populations, coherences, or fidelity. We therefore separate two issues that should not be conflated: a) closure of the mean kinetic-OAM dynamics, and b) transport of the underlying vortex state.

\section{Failure of the BMT-like closure for the mean kinetic OAM}

We first show directly that the passage from Eq.~(8) to Eq.~(9) in
Ref.~\cite{Karlovets} is not justified for the \emph{mean kinetic} OAM.
The obstruction is already visible in the authors' own formulas:
their Eq.~(8) expresses $\langle L_z\rangle$ through the transverse second
moment $\langle \rho^2\rangle(\tau)$, whereas Eq.~(9) is then promoted to a
closed BMT-like first-moment precession law for $\langle \mathbf L\rangle$.
For a uniform longitudinal field these two statements are incompatible, except
in the nongeneric matched case in which the breathing amplitude vanishes
identically.

Let the external magnetic field be homogeneous and longitudinal,
\begin{equation}
    \mathbf H_0 = H_0 \hat{\mathbf z},
\end{equation}
and define
\begin{equation}
    \Omega = \frac{|e|H_0}{2m} > 0,
    \qquad
    \omega_c = 2\Omega .
\end{equation}
According to Eq.~(8) of Ref.~\cite{Karlovets}, the mean kinetic OAM satisfies
\begin{equation}
    \langle L_z\rangle(\tau)
    =
    \langle L_z^{(c)}\rangle
    + \frac{eH_0}{2}\,\langle \rho^2\rangle(\tau).
    \label{eq:Lz_from_rho2}
\end{equation}
For a centered vortex mode with definite canonical OAM projection,
\begin{equation}
    \langle L_z^{(c)}\rangle = l ,
\end{equation}
since $L_z^{(c)}$ commutes with the homogeneous-field Hamiltonian.

To make the contradiction explicit, it is enough to consider one exact family,
namely centered, cylindrically symmetric, self-similar breathing Landau/LG packets
(squeezed Landau states). After removal of the Larmor rotation, the
transverse dynamics reduces to the 2D isotropic harmonic oscillator with
frequency $\Omega$, and the exact scaling parameter $b(\tau)$ obeys
\begin{equation}
    \ddot b + \Omega^2 b = \frac{\Omega^2}{b^3}.
\end{equation}
With
\begin{equation}
    b(0)=b_0,
    \qquad
    \dot b(0)=0,
\end{equation}
the exact solution is
\begin{equation}
    b^2(\tau)
    =
    \frac12\left[b_0^2+\frac{1}{b_0^{2}}\right]
    +
    \frac12\left[b_0^2-\frac{1}{b_0^{2}}\right]\cos(\omega_c\tau).
    \label{eq:b2_exact}
\end{equation}
Hence the breathing oscillates exactly at the cyclotron frequency.

For this family,
\begin{align}
   & \langle \rho^2\rangle(\tau)
    =
    \frac{\rho_H^2}{2}\,N\,b^2(\tau), \nonumber \\
    &N=2n_r+|l|+1, \\
   & \rho_H=\frac{2}{\sqrt{|e|H_0}}. \nonumber
\end{align}

Therefore
\begin{align}
\langle L_z\rangle(\tau)
    =
    l+\sigma N\,b^2(\tau),~~   
    \sigma=\frac{eH_0\rho_H^2}{4}=\operatorname{sgn}(eH_0).
    \label{eq:Lz_shell_form}
\end{align}
Equation above is fully consistent with Ref.\cite{Greenshields}.

To avoid any misleading dependence on dimensional scales, introduce the
dimensionless cyclotron time
\begin{equation}
    t=\omega_c\tau,
\end{equation}
and the shell-normalized mean kinetic OAM
\begin{equation}
    \ell_z(t)=\frac{\langle L_z\rangle(\tau)}{N}.
\end{equation}
For definiteness, take
\begin{equation}
    eH_0>0,\qquad l>0,\qquad n_r=0,\qquad l\gg1,
\end{equation}
so that $N=l+1$ and $l/N=1+\mathcal{O}(l^{-1})$.
Using Eq.~\eqref{eq:b2_exact}, one obtains
\begin{align}
    \ell_z(t)
    =
    1
    +\frac12\left[b_0^2+\frac{1}{b_0^2}\right]
    +\frac12\left[b_0^2-\frac{1}{b_0^2}\right]\cos t
    +\mathcal{O}\!\left(\frac1l\right),
    \label{eq:ellz_dimensionless}
\end{align}
and therefore
\begin{equation}
    \frac{d\ell_z}{dt}
    =
    -\frac12\left[b_0^2-\frac{1}{b_0^2}\right]\sin t
    +\mathcal{O}\!\left(\frac1l\right).
    \label{eq:dellz_dt}
\end{equation}
This derivative is nonzero for every mismatched packet $b_0\neq1$, and is of
order unity for generic finite mismatch independent of $l$.

It is not suppressed by any factor of $1/l$, $1/N$, or by the bare Landau scale
$\rho_H$.
In particular, the oscillatory part already exceeds the normalized canonical
contribution (which tends to $1$ as $l\to\infty$) once
\begin{equation}
    \frac12\left[b_0^2-\frac{1}{b_0^{2}}\right]>1,
\end{equation}
that is,
\begin{equation}
    b_0>\sqrt{1+\sqrt2}\approx1.55.
\end{equation}
Now rewrite the spinless version of Eq.~(9) of Ref.~\cite{Karlovets} in the same dimensionless time
\begin{equation}
    \frac{d\langle \mathbf L\rangle}{dt}
    =
    \frac12\,\hat{\mathbf z}\times \langle \mathbf L\rangle .
    \label{eq:BMT_like_dimless}
\end{equation}
Therefore its longitudinal component is identically conserved,
\begin{equation}
    \frac{d\ell_z}{dt}=0.
    \label{eq:BMT_like_dimless_z}
\end{equation}
Equations~\eqref{eq:dellz_dt} and \eqref{eq:BMT_like_dimless_z} are incompatible unless
$b_0=1$.
Thus the failure of the closure is not a matter of a parametrically small correction.
Even in dimensionless shell-normalized variables, Eq.~(9) of Ref.~\cite{Karlovets}
predicts exact constancy, whereas the exact breathing family exhibits a nonzero
oscillation, and an $\mathcal{O}(1)$ oscillation for generic finite mismatch.

\section{On the Appendix A small-correlation assumption.}

The previous argument already provides an explicit counterexample to the claimed closure. We now show something stronger. Namely, the Appendix A small-correlation assumption suppresses the transverse kinetic OAM itself.

In a homogeneous longitudinal field with the symmetric gauge
\(
A_x=-H_0 y/2,\;
A_y=H_0 x/2,\;
A_z=0
\),
one has
\(
p_x=p_x^{(c)}+eH_0 y/2
\),
\(
p_y=p_y^{(c)}-eH_0 x/2
\),
and
\(
p_z=p_z^{(c)}
\).
Therefore
\begin{equation}
L_x
=
y p_z-z p_y
=
y p_z^{(c)}-z p_y^{(c)}+\frac{eH_0}{2}xz,
\label{eq:Lx_kinetic_mixed}
\end{equation}
and
\begin{equation}
L_y
=
z p_x-x p_z
=
z p_x^{(c)}-x p_z^{(c)}+\frac{eH_0}{2}yz.
\label{eq:Ly_kinetic_mixed}
\end{equation}
If the Appendix A argument is interpreted as asserting that each of the mixed correlators entering Eq.~(A5) is negligible in the quasi-classical closure, then one should impose the same type of smallness assumption used after Eq.~(A5), namely
\begin{equation}
|\langle y p_z^{(c)}\rangle|,\ |\langle z p_y^{(c)}\rangle|,\ 
|\langle z p_x^{(c)}\rangle|,\ |\langle x p_z^{(c)}\rangle|
\le \varepsilon ,
\end{equation}
together with
\begin{equation}
\left|\frac{eH_0}{2}\langle xz\rangle\right|,\ 
\left|\frac{eH_0}{2}\langle yz\rangle\right|
\le \varepsilon .
\end{equation}
If, instead, only a cancellation of their sum were assumed, then Appendix~A would not
provide a controlled term-by-term justification for dropping Eq.~(A4) in the first place.

By the triangle inequality,
\begin{equation}
|\langle L_x\rangle|
\le
|\langle y p_z^{(c)}\rangle|
+
|\langle z p_y^{(c)}\rangle|
+
\left|\frac{eH_0}{2}\langle xz\rangle\right|
\le 3\varepsilon,
\end{equation}
and similarly
\begin{equation}
|\langle L_y\rangle|
\le
|\langle z p_x^{(c)}\rangle|
+
|\langle x p_z^{(c)}\rangle|
+
\left|\frac{eH_0}{2}\langle yz\rangle\right|
\le 3\varepsilon.
\end{equation}
Hence, in the strict closure limit $\varepsilon\to0$,
\begin{equation}
\langle L_x\rangle=\langle L_y\rangle=0.
\end{equation}
But Eq.~(9) is precisely supposed to describe precession of the mean kinetic-OAM vector.
Therefore the approximation used to suppress the omitted term in Eq.~(A4) simultaneously
suppresses the transverse kinetic OAM itself.
The neglected mixed correlators are thus not a small quantum correction, they are the very
building blocks of the transverse kinetic OAM whose evolution Eq.~(9) of the Ref.~\cite{Karlovets} is meant to capture.

For instance, if
\(
\langle \mathbf L\rangle \sim l(\sin\theta,0,\cos\theta)
\)
with fixed $\theta\neq0$, then
\[
l|\sin\theta|
\lesssim
|\langle y p_z^{(c)}\rangle|
+
|\langle z p_y^{(c)}\rangle|
+
\left|\frac{eH_0}{2}\langle xz\rangle\right|.
\]
Hence
\[
\max\!\left\{
|\langle y p_z^{(c)}\rangle|,
|\langle z p_y^{(c)}\rangle|,
\left|\frac{eH_0}{2}\langle xz\rangle\right|
\right\}
\gtrsim \frac{l|\sin\theta|}{3},
\]
so at least one of the supposedly small mixed correlators must actually scale as
$\mathcal{O}(l)$.


\section{Mean kinetic OAM is not a transport equation for a vortex state}

Even if one grants, for the sake of argument, that a closed evolution equation for the
mean kinetic OAM could be derived, this would still not justify the state-level claims made
in Ref.~\cite{Karlovets}.
A transport law for a few low-order moments is not, in general, a transport law for the
underlying vortex quantum state.

A quantum state is not determined by the expectation values of a small set of observables;
rather, one needs the wavefunction or density operator, or equivalently an informationally
complete set of observables~\cite{LandauQM,Messiah,Sakurai}.
Accordingly, the map
\begin{equation}
\rho \mapsto \langle \hat{\mathbf L}\rangle = {\rm Tr}(\rho\,\hat{\mathbf L})
\end{equation}
is highly non-injective.
Distinct density operators may have the same $\langle \hat{\mathbf L}\rangle$ while differing
in their OAM spectra, inter-mode coherence, and fidelity to an initially prepared vortex mode.
Therefore an equation for $\langle \hat{\mathbf L}\rangle$ is not a state-transport equation.
It is an Ehrenfest-type equation for a low-order moment.

This is precisely where the spin analogy breaks down.
For a two-level spin system, the Bloch vector determines the full $2\times2$ density matrix.
No analogous state-completeness holds for a generic OAM wavepacket. Unlike the spin-1/2 case, a few first moments do not determine the full density operator \cite{LandauQM,Messiah,Sakurai}. Moreover, OAM states generally occupy a high-dimensional mode space rather than a two-level manifold \cite{Calvo}.
Hence terms such as polarization, depolarization and spin-like resonance do not
automatically acquire the same meaning for OAM.
Such language becomes justified only after one identifies a closed reduced mode manifold and
demonstrates that the transport remains confined to it.

The logical failure can be exhibited explicitly.
Fix a reference axis $\mathbf n$ and define
\begin{equation}
\hat L_{\mathbf n}\equiv \hat{\mathbf L}\!\cdot\!\mathbf n,
\qquad
\hat L_{\mathbf n}|\ell\rangle=\ell|\ell\rangle,
\end{equation}
suppressing additional mode labels.
Let an initially pure eigenmode $|\ell_0\rangle$ pass first through a weak symmetry-breaking
interface represented by a unitary $U_{\rm int}$ that couples $\Delta\ell=\pm2$, and then through
a section symmetric about the same axis,
\begin{equation}
U_{\rm sym}(\mu)=e^{-i\mu\hat L_{\mathbf n}/\hbar}.
\end{equation}
To order $\varepsilon^2$,
\begin{align}
&|\psi(\mu)\rangle
\equiv U_{\rm sym}(\mu)U_{\rm int}|\ell_0\rangle \\ \nonumber
&\simeq e^{-i\ell_0\mu}\Big[(1-\varepsilon^2)|\ell_0\rangle
+\varepsilon e^{-i2\mu}|\ell_0+2\rangle
+\varepsilon e^{+i2\mu}|\ell_0-2\rangle\Big],
\label{eq:toy_state_transport}
\end{align}
with $\varepsilon\ll1$.
Then
\begin{equation}
\langle\psi(\mu)|\hat L_{\mathbf n}|\psi(\mu)\rangle
=\ell_0+\mathcal{O}(\varepsilon^4),
\end{equation}
so the mean OAM projection is preserved to this order.
But the state is not preserved:
\begin{equation}
|\langle\ell_0|\psi(\mu)\rangle|^2\simeq1-2\varepsilon^2,
\end{equation}
and the sideband coherence acquires the relative phase $e^{-i4\mu}$.
Thus stability of the mean OAM does not imply preservation of mode purity, coherence,
or fidelity.

Therefore, even a genuinely closed and internally consistent equation for the mean kinetic OAM
would still be insufficient to support claims about preservation, depolarization, or
controlled manipulation of twisted quantum states. Such claims require a transport theory for the mode-resolved density matrix
\begin{equation}
\rho_{\ell\ell'}(s),
\end{equation}
or equivalently for the populations $P(\ell)$, inter-mode coherence, and fidelity.
Without that state-resolved dynamics, one has at most a reduced theory of low-order moments,
not a theory of vortex-state transport.

\bibliographystyle{apsrev4-2} 
\bibliography{refs}  

\end{document}